\documentclass[prl,twocolumn,showpacs]{revtex4}
\usepackage{graphics}
\usepackage{graphicx}
\usepackage{amsmath}
\usepackage{amstext}
\usepackage{xcolor}
\usepackage{amssymb}
\usepackage{hyperref}

\begin{document}

\title{Hall Effect in spinor condensates} \author{Mathieu
  Taillefumier$^1$, Eskil K. Dahl$^1$, Arne Brataas$^1$ and Walter
  Hofstetter$^2$}\affiliation{$^1$Department of Physics, Norwegian
  University of Science and Technology, N-7491 Trondheim,
  Norway\\
  $^2$ Institut f\"ur Theoretische Physik, J. W. Goethe-Universit\"at
  Max-von-Laue-Str. 1, 60438 Frankfurt/Main, Germany}

\date{\today }
\begin{abstract}
  We consider a neutral spinor condensate moving in a periodic
  magnetic field. The spatially dependent magnetic field induces an
  effective spin dependent Lorentz force which in turn gives rise to a spin
  dependent Hall effect. Simulations of the Gross-Pitaevskii equation
  quantify the Hall effect. We discuss possible experimental realizations.
\end{abstract}
\maketitle

In 1879, Edwin Hall discovered that charges in a thin metallic plate
exposed to a perpendicular magnetic field accumulate transverse to
applied electric and magnetic fields. This Hall effect is used
intensively to characterize metals and semi-conductors. Modern
interest in Hall effects started with the experimental discovery of
the quantum Hall effect in two-dimensional electrons gases (2DEG) and
the seemingly disconnected notion of Berry curvature~\cite{Berry1984}
and non abelian gauge field~\cite{Wilczek1984}. Currently these
notions explain the integer and fractional quantum Hall effect in
2DEG, the anomalous Hall effect (AHE) in ferromagnetic metals or
semi-conductors and the spin analogue known as spin Hall
effect~\cite{Sinova2004}.

Three mechanisms contribute to the AHE. Two of them, the
side-jump~\cite{Berger1970} and the skew scattering~\cite{Smit1955},
result from carrier scattering off impurities while the
Karplus-Luttinger term~\cite{Karplus1954}, also known as Berry
curvature contribution, is related to spin-orbit coupling or non
trivial magnetic order~\cite{Onoda2003b} in the band structure. The
latter contribution has attracted considerable theoretical attention
although it is very difficult to study experimentally in ferromagnetic
compounds because of the additional contributions from impurities or
defects.

The realization of Bose-Einstein condensation in magnetic or optical
trap opens up the possibility of research at the border between atomic
and condensed matter physics. Analogies between condensates in
rotating magnetic traps and electrons in strong magnetic fields allow
studies of the integer and fractional quantum Hall
states~\cite{Bhat2007,Viefers2008}. Recently, the realization of spin
dependent optical lattices in combination with cold atomic
gases~\cite{Mandel2003} offers the possibility to study spin-dependent
transport phenomena. The neutrality of these gases, {\it i.e.} the
absence of a classical Lorentz force and the absence of impurity
scattering, allows us to study the spin Hall
effect~\cite{Liu2007,Bliokh2006,Zhu2006,Dudarev2004,Ruseckas2005} and
the Berry curvature contribution of the AHE~\cite{Dudarev2004} in a
clean and controllable environment. So far, the single particle
approximation is often used to explain the AHE and SHE in cold atomic
gases~\cite{Dudarev2004,Liu2007,Zhu2006}. Although the single particle
approximation can be used to describe qualitatively the AHE in
multicomponent condensates, two-body interactions should be included
in the theory for a quantitative description of the Hall effect.

Magnetic microtraps~\cite{Fortagh2007} might also be used to study the
AHE and SHE in Bose condensates. In this case, the Zeeman effect leads
to a coupling between the spin of the atoms and the magnetic field of
the microtrap.  Various geometries of magnetic field distributions
including one-dimensional magnetic lattices have been explored
experimentally~\cite{Fortagh2007}, and two-dimensional magnetic
lattices can be created using array of magnetic
cylinders~\cite{Nielsch2001} or magnetic dots for instance.

In this letter, we investigate the Hall effect in spinor condensates
where the magnetic field is created by a two-dimensional array of
magnetic cylinders.  The motion of the condensate through the magnetic
lattice generates a non abelian gauge field that acts on the
condensate as an effective spin dependent Lorentz
force~\cite{Aharonov1992}. Similarly to the case of
fermions~\cite{Bruno2004}, we first study the case where the two-body
interactions are negligible. Then we study numerically the effect of
the velocity of the condensate and of the non linear interactions on
the Hall Effect.  Finally we propose an experimental setup where this
effect can be observed.

Let us consider a quasi-two-dimensional spinor condensate of spin $F$
moving in a magnetic field ${\bf B}({\bf r})$ created by a 2D array of
magnetic cylinders. The Hamiltonian describing this system
is~\cite{Ohmi1998,Ho1998,Castin1999}
\begin{eqnarray}
  \label{eq:1}
  \mathcal{H}&=&\int d^2{\bf r} \psi^\dagger_\alpha({\bf r})\left( 
    \left[-\frac{\hbar^2}{2 M} \nabla_{\bf r}^2+V({\bf
        r})\right]\delta_{\alpha\delta}+({\bf B}({\bf r})\cdot {\bf F})_{\alpha\delta}\right.\nonumber\\ 
  &+& \left.\sum^{F}_{s=0} \frac{\sqrt{2\pi}\hbar^2a_{2s}}{M a_z} (\mathcal{P}^{2 s})_{\alpha\beta\gamma\delta}
    \psi^\dagger_\beta({\bf r})\psi_\gamma ({\bf
      r})\right)\psi_\delta({\bf r})
\end{eqnarray}
%
where $\psi_\alpha({\bf r})$ with $\alpha=-F,\ldots,F$ are the spin
components of the quantum field operator $\Psi({\bf r})$, and
$\mathcal{P}^{2s}$ is the projection operator that projects the spin
state of the atoms pair into the total spin state $2s$ . $F^{l=x,y,z}$
are matrices of spin $F$ and the constants $a_{2s}$ are the s-wave
scattering lengths of the total spin channel $2s=0,2,\ldots 2F$. The
confinement in the $z$ direction appears in the Hamiltonian via the
oscillator length $a_z=\sqrt{\hbar/M\omega_z}$, $\omega_z=2\pi f_z$
where $f_z$ is the frequency of the trap along the $z$ direction and
$M$ the mass of the atoms. The scalar potential $V({\bf r})$ describes
any external trap potential. Following Ref.~\cite{Bruno2004}, we apply
a rotation $\mathcal{T}({\bf r})$ of the quantization axis along the
direction ${\bf n}({\bf r})$ of the magnetic field ${\bf B}({\bf r})$
to (\ref{eq:1}). The transformed Hamiltonian becomes
\begin{eqnarray}
  \label{eq:2}
  \mathcal{H}_{\mathcal{T}}&=&\int d^2{\bf r} \psi^\dagger_\alpha ({\bf r})\left(
    \left[-\frac{\hbar^2}{2 M} \left(\nabla_{\bf r}-i {\bf A}_g({\bf r})\right)^2+V({\bf
        r})\delta_{\alpha\gamma}\right]\right.\nonumber\\ 
  &+&\left. |{\bf B}({\bf r})| F_z^{\alpha\delta}\right. \nonumber\\ &+&\left.\sum^{F}_{s=0}
    \frac{\sqrt{2\pi}\hbar^2a_{2s}}{M a_z} 
    (\mathcal{P}^{2 s})_{\alpha\beta\gamma\delta}
    \psi^\dagger_\beta({\bf r})\psi_\gamma ({\bf
      r})\right)\psi_\delta({\bf r}),
\end{eqnarray}
where ${\bf A}_g({\bf r})=-i\mathcal{T}^\dagger({\bf r})\nabla_{\bf r}
\mathcal{T}({\bf r})={\bf A}^i_g({\bf r}) F^i$ with $i=x,y,z$ is the
spin dependent gauge field resulting from the rotation of the spin axis. 
Explicit expressions for ${\bf A}^i_g({\bf r})$ ($i=x,y,z$) can be found in
Ref.~\cite{Bohm2003}. The two-body interactions are not affected by the
gauge transformation because they are short ranged and the projection
operators are invariant against any rotation of the spin axis of both
particles. Excluding the two-body interactions, Eq.~(\ref{eq:2}) is a
generalization of Eq.~(2) of Ref.~\cite{Bruno2004} for both fermions and
bosons.
The present work has two additional ingredients, the non linear
interactions and the relaxation of the adiabaticity condition. The non
abelian gauge field ${\bf A}_g({\bf r})$ contains terms that are
proportional to $F^x$ and $F^y$ that induce transitions between the
different spin-polarized states. 

{\it Approximate analytical calculations:} Let us initially assume
that the spin-flip transitions can be neglected, {\it i.e}., that the
spin adiabatically follows the magnetic field direction ${\bf n}({\bf
  r})$. This approximation will be lifted later on. In contrast to
$F=1/2$, the expansion of Eq.~(\ref{eq:2}) gives rise to two spin-flip
terms that are proportional to ${\bf A}_g({\bf r})$ and ${\bf
  A}_g({\bf r})\cdot {\bf A}_g({\bf r})$. The adiabatic approximation
is valid when $\nu_1=\frac{\hbar v}{2 \xi\Delta_z}\ll 1$ and
$\nu_2=\frac{\hbar^2}{2 m \xi^2\Delta_z}\ll 1$ are fulfilled. The
constant $v$ is the group velocity of the particles, $\xi$ is a
characteristic length of variations of ${\bf n}({\bf r})$ and
$\Delta_z$ is the Zeeman splitting. The two-body interactions are
included in Eq.~\ref{eq:2} through a spin-spin interaction term that can be
decomposed into a scalar and a spin-dependent
contribution~\cite{Ho1998}.
The typical energy scale of the scalar inter-atomic interaction
component is of the order of magnitude of the Zeeman splitting. 
The spin-dependent part of the two-body interactions, on the other hand,
 can be neglected because it is two to three orders of magnitude
smaller than the scalar contribution. For this semi-classical
analysis, we neglect the two-body interactions and assume that the
spinor condensate is in a spin polarized state $m\in
\left[-F,\ldots,F\right]$. The effective Hamiltonian can be written as
\begin{eqnarray}
  \label{eq:3}
  \mathcal{H}_{\text{eff}}&=&\int d^3{\bf r} \psi^\dagger_m ({\bf r})
  \left[\frac{\hbar^2}{2 M} \left(-i\nabla_{\bf r}+ m {\bf A}_g^z({\bf
        r})\right)^2
    +V({\bf r})\right.\nonumber\\
  &+&\left. m |{\bf B}({\bf r})|\right]\psi_m({\bf r}).
\end{eqnarray}
There is a difference between bosons and fermions. For fermions, all
spin polarized states are affected by the gauge field ${\bf
  A}^z_g({\bf r})$ while for bosons, the spin polarized state $m=0$ is
not affected by the magnetic texture. So for $m\neq 0$ the field ${\bf
  b}_g({\bf r})=\nabla \times {\bf A}^z_g({\bf
  r})=\varepsilon_{ijk}n_i\partial_xn_j\partial_y n_k$ acts on the
spin polarized state as an ordinary magnetic field on a fictitious
charge $q=m e$ ($-e$ is the electron charge), and then gives rise to
an effective Lorentz force~\cite{Aharonov1992} that can induce a Hall
effect.

For a qualitative analysis, we consider the case where the average
value of the field $\left<{\bf b}_g({\bf r})\right>_s$ over one unit
cell acts on the dynamics of the condensate. Firstly, the average
value $\left<{\bf b}_g({\bf r})\right>_s=4\pi n/S$ with
($n\in\mathbb{Z}$) and $S$ the surface of the unit cell, is quantized
and depends only on few geometrical parameters such as the type and
the period of the lattice.  This value can be modified by applying a
small constant magnetic field on top of the magnetic
distribution~\cite{Bruno2004}. Secondly, the spatial dependence of the
field ${\bf b}_g({\bf r})$ and the potential $|{\bf B}({\bf r})|$ are
treated phenomenologically as causing momentum scattering that is
characterized by the relaxation time $\tau$. We consider moreover that
the condensate is under the influence of a force ${\bf F}=F_0 {\bf x}$
where $F_0$ is the amplitude. Such forces can be induced by gravity
for instance.
The dynamics of the center of mass of the spin polarized state is
described by:
\begin{equation}
  \label{eq:4}
  M \frac{d \left<{\bf v}\right>}{d t}= F_0 {\bf x} 
  - m\left(\frac{\left<{\bf v}\right>}{\tau}
    + \hbar\left<{\bf v}\right>\times \left<{\bf b}_g\right>_s\right),
\end{equation}
where $\left<{\bf v}\right>$ is the center-of-mass velocity of the
polarized state. The solutions of Eq.~(\ref{eq:4}) are simple and the
wave packet trajectories go from closed orbits when there is no
scattering and no force to open orbits when scattering and/or external
force are present. One can also show that the quantity
$\sigma_{xy}=\left< v_y\right>/F_0$ is similar to the Drude formula of
the Hall conductivity component. The Hall effect is stronger when the
 spin $F$ increases or the surface of the unit cell decreases.

\begin{figure}
  \centering
  \includegraphics[scale=1]{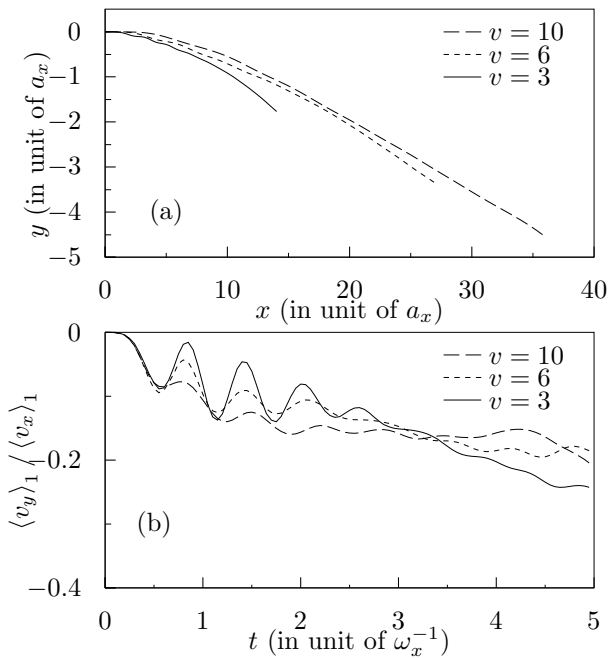}
  \caption{(a) Transverse position as a function of longitudinal
    coordinate of the center of mass of a $F=1$ condensate for
    $t\omega_x\leq 5$ and different values of the initial group
    velocity $v$ (in units of $a_x\omega_x$). (b) Time dependence of
    the ratio of average transverse and longitudinal velocity. The
    initial state is prepared in the ferromagnetic state $m=1$.  }
  \label{fig:1}
\end{figure}

\begin{figure}[t]
   \centering
   \includegraphics[scale=1]{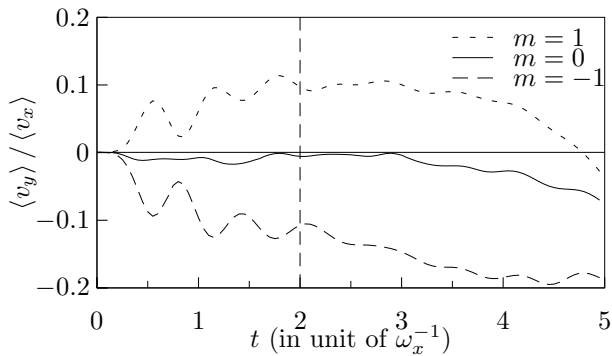}
   \caption{Time dependence of the ratio
     $\left<v_y\right>/\left<v_x\right>$ calculated for different
     initial polarizations and fixed initial velocity $v=6$. 
     The vertical lines indicates the separation between the adiabatic regime
     where spin flip contributions are negligible from the regime
     where they dominate the dynamics of the condensate.}
   \label{fig:2}
 \end{figure}

 {\it Numerical simulations}: A quantitative study of the Hall effect
 in this systems can be done by solving Eq.~(\ref{eq:1}) numerically
 using mean field theory. The order parameters of the spinor
 condensate, which are the expectation values of the field operators
 $\psi_\alpha({\bf r})$, obey the time dependent Gross-Pitaevskii
 equation (TDGP) obtained from Eq.~(\ref{eq:1}) by substituting
 $\psi_\alpha({\bf r})$ by its average. We solve the TDGP
 equation using an explicit Runge-Kutta method of order six coupled to
 the Fast Fourier transform for the spatial derivatives. For the
 magnetic field distribution, we consider a square lattice of
 magnetic cylinders. 

 The initial state of the system is obtained in the following way. We
 first calculate the ground state of a spinor condensate of spin $F=1$
 composed of $2000$ atoms of rubidium~\cite{note} in an harmonic trap
 with $\omega_x=\omega_y=2 \pi\times 50\ $Hz ($\omega_z=2\pi \times
 1000\ $Hz). Then the condensate is shifted to one side of the trap
 and the wave-function is multiplied by a phase factor $\exp(i v x)$
 where $v$ is the group velocity. At $t=0$ the condensate is released
 from the harmonic trap and is able to move freely through the
 magnetic lattice.

 To characterize the Hall effect, we calculate the ratio between the
 transverse and longitudinal velocity
 \begin{equation}
  \label{eq:5}
  \left<v_y\right>_m/\left<v_x\right>_m=\sigma_{xy}/\sigma_{xx},
\end{equation}
where $m$ indicates the initial polarization of the condensate at
$t=0$ (in the global quantization axis which is along the $z$ axis)
and $\left<v_{i=x,y}\right>_m$ is the average value of the velocity
operator. $\sigma_{xy}$ and $\sigma_{xx}$ are the transverse and
longitudinal "conductivities".  This ratio corresponds physically to
the tangential of the angle between the electrical field and the
velocity in the case of electronic systems. For convenience, we
express all quantities in reduced units $x\to a_x x$ with
$a_x=\sqrt{\frac{\hbar}{M\omega_x}}\approx 1.59$ $\mu$m and the energy
$\varepsilon \to \varepsilon\hbar\omega_x$.  Finally we choose the
following set of parameters of the magnetic field. The period of ${\bf
  B}({\bf r})$ is $p=7.5\ a_x$, the amplitude is $B=10 \hbar\omega_x$
and the distance separating the condensate from the surface of the
magnetic array is $h=1.5\ a_x$.

The dynamical behavior of the TDGP equation is visualized in
Fig.~\ref{fig:1} which represents the ballistic trajectory of the
center-of-mass of the Bose condensate (fig.~\ref{fig:1}-a) and the
Hall angle (fig.~\ref{fig:1}-b) for different values of the initial
velocity. Fig.~\ref{fig:1}-a shows that the trajectory of the
condensate prepared in the state $m=1$ is curved in the $y$ direction
because of the action of the gauge field on the dynamics of the
condensate as qualitatively described by
Eq.~\ref{eq:3}. Fig.~\ref{fig:1}-b represents the time dependence of
the ratio $\left<v_y\right>_{m=1}/\left<v_x\right>_{m=1}$.  It shows
pronounced oscillations at early stages of the evolution that can be
understood qualitatively by considering the sign of the effective
Lorentz force that is acting on the condensate. Since the size of the
condensate is smaller than the size of the unit cell, part of the
condensate will experience a positive Lorentz force while its
complementary part will experience a negative Lorentz force. Due to
the distribution of particles in the condensate and the different
topology of the regions where the field ${\bf b}_g({\bf r})$ is
positive or negative, the Lorentz force changes sign when the
condensate is moving through the lattice and the $y$ component of the
group velocity oscillates with time. The oscillations of
$\left<v_y\right>_{m=1}/\left<v_x\right>_{m=1}$ quickly disappear when
$t$ increases because (i) the condensate expands and (ii) the
magnetization is not conserved so the mean field state describing the
condensate tends to a state with equipartition in populations ($n_{\pm
  1,0}=1/3$).
Numerical simulations show that these oscillations remain when the
size of the unit cell is smaller than the size of the condensate, as it
should be expected.

The dynamical properties of the system can be separated into two
different regimes; the adiabatic regime where the spin polarized state
follows adiabatically ${\bf B}({\bf r})$ as described by our
approximate analytical calculations and a second regime that is
dominated by the spin flip terms. The adiabatic regime can be
identified at early time of the evolution of the condensate by
calculating the ratio $\left<v_y\right>/\left<v_x\right>$ for the spin
polarized states $m=1$ and $m=-1$. In this case, Eq.~(\ref{eq:3})
shows that the two angles should differ by a sign which can be seen on
the left part of Fig.~\ref{fig:2}. This behavior is not peculiar to
Bose condensates and should be observable with fermionic atomic gases
or electrons gases provided that the gas is initially fully polarized
and there is no disorder. As expected from Eq.~(\ref{eq:3}), there
should be no Hall effect for the polar state ($m=0$) which can also be
observed in Fig.~\ref{fig:2}. Fig.~\ref{fig:2} shows that
$\left<v_y\right>_0/\left<v_x\right>_0$ of an initial polar state
fluctuates around zero. These fluctuations are not affected by the
spin-dependent part of the two-body interactions but are related to
the spin-flip contributions of the gauge field ${\bf A}_g({\bf
  r})$. At longer time (right part Fig.\ref{fig:2}), the dynamics of
the condensate is governed by the spin-flip terms and the description
of the Hall effect in terms of the adiabatic approximation in
(\ref{eq:3}) is not valid anymore.

\begin{figure}
  \includegraphics[scale=1]{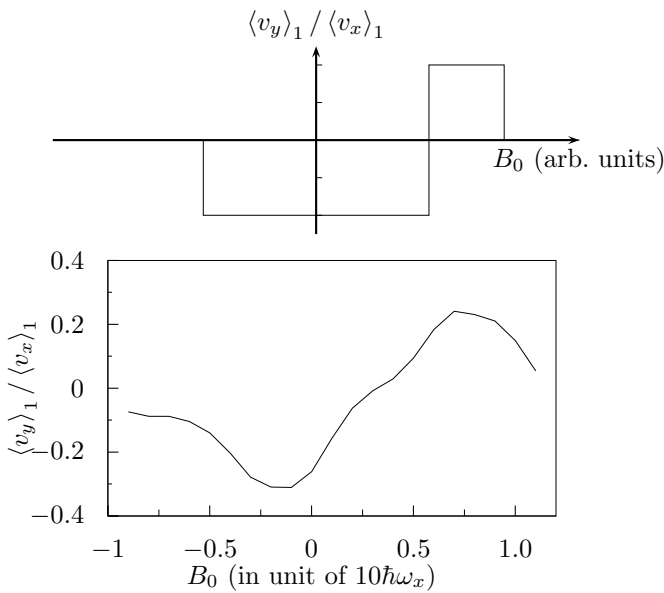}
  \caption{(upper panel) schematic dependence of the Hall conductivity
    when a constant magnetic field is applied on the top of the
    magnetic lattice. The abrupt changes in
    $\left<v_y\right>_1/\left<v_x\right>_1$ are related to the flux of
    ${\bf b}_g({\bf r})$ which changes sign at a peculiar value of
    $B_0$. (lower panel) numerical results for the ratio
    $\left<v_y\right>_1/\left<v_x\right>_1(5\omega_x^{-1})$ as a
    function of the magnetic field amplitude.  }
    \label{fig:3}
\end{figure}

Finally, we study the influence of a small constant magnetic field of
amplitude $B_0$ applied on top of the magnetic lattice in the
direction perpendicular to the plane of motion. Contrary to
Ref. \cite{Bruno2004} which consider the electronic case, there is no
classical Hall effect because the atoms are neutral. 
Therefore the features of the anomalous Hall effect are
easily observed in Bose condensates. In the semiclassical limit, {\it
  i.e.} when the spatial dependence of the gauge field is neglected,
the variations of the Hall angle are abrupt because the average value
of ${\bf b}_g({\bf r})$ can only change at peculiar values of $B_0$
which depends on the lattice. Such variations are schematically
represented on the upper panel of Fig.\ref{fig:3}. The general
behavior described in the upper panel of Fig.\ref{fig:3} can be
reproduced numerically using the full time-dependent GP equation but the variations of the ratio
$\left<v_y\right>/\left<v_x\right>$ are continuous rather than
abrupt (lower panel of Fig.\ref{fig:3}).  
The numerical simulations also show that the Hall effect
decrease rapidly when the amplitude of the constant magnetic field
increases.

To experimentally verify this theory, we propose to image the motion
of a spin polarized condensate during its evolution 
in a microtrap with an additional periodic 2d magnetic field 
and deduce the ratio $\left<v_y\right>/\left<v_x\right>$ by
direct measurement of the velocity. The magnetic lattice can be
created using arrays of magnetic cylinders or by patterning lattices of 
interconnected current loops. The condensate
can be accelerated through gravity by tilting the microtrap.

To conclude, we have shown that spinor condensates in a magnetic
lattice can be used to explore the geometrical phase contribution of
the anomalous Hall effect. We found that this contribution is
characterized by an adiabatic regime which is valid in the short time
limit and a long time regime where spin-flip processes dominate the
condensate dynamics. We also demonstrated that the AHE is strongly
affected by an additional constant external magnetic field.

The authors thank M. Snoek, B. Canals, C. Lacroix, V. Dugaev, P. Bruno
and J. Dalibard for discussion.  This work was supported in part by
Forschergruppe FOR 801 of the Deutsche
Forschungsgemeinschaft.


\begin{thebibliography}{26}
\expandafter\ifx\csname natexlab\endcsname\relax\def\natexlab#1{#1}\fi
\expandafter\ifx\csname bibnamefont\endcsname\relax
  \def\bibnamefont#1{#1}\fi
\expandafter\ifx\csname bibfnamefont\endcsname\relax
  \def\bibfnamefont#1{#1}\fi
\expandafter\ifx\csname citenamefont\endcsname\relax
  \def\citenamefont#1{#1}\fi
\expandafter\ifx\csname url\endcsname\relax
  \def\url#1{\texttt{#1}}\fi
\expandafter\ifx\csname urlprefix\endcsname\relax\def\urlprefix{URL }\fi
\providecommand{\bibinfo}[2]{#2}
\providecommand{\eprint}[2][]{\url{#2}}

\bibitem[{\citenamefont{Berry}(1984)}]{Berry1984}
\bibinfo{author}{\bibfnamefont{M.~V.} \bibnamefont{Berry}},
  \bibinfo{journal}{Proc. R. Soc.} \textbf{\bibinfo{volume}{392}},
  \bibinfo{pages}{45} (\bibinfo{year}{1984}).

\bibitem[{\citenamefont{Wilczek and Zee}(1984)}]{Wilczek1984}
\bibinfo{author}{\bibfnamefont{F.}~\bibnamefont{Wilczek}} \bibnamefont{and}
  \bibinfo{author}{\bibfnamefont{A.}~\bibnamefont{Zee}},
  \bibinfo{journal}{Phys.\ Rev.\ Lett.} \textbf{\bibinfo{volume}{52}},
  \bibinfo{pages}{2111} (\bibinfo{year}{1984}).
\bibitem[{\citenamefont{Sinova et~al.}(2004)\citenamefont{Sinova, Culcer, Niu,
  Sinitsym, Jungwirth, and MacDonald}}]{Sinova2004}
\bibinfo{author}{\bibfnamefont{J.}~\bibnamefont{Sinova}~{\it et al.}},
 \bibinfo{journal}{Phys. Rev. Lett.}
  \textbf{\bibinfo{volume}{92}}, \bibinfo{pages}{126603}
  (\bibinfo{year}{2004}).

\bibitem[{\citenamefont{Berger}(1970)}]{Berger1970}
\bibinfo{author}{\bibfnamefont{L.}~\bibnamefont{Berger}},
  \bibinfo{journal}{Phys.\ Rev.\ B} \textbf{\bibinfo{volume}{2}},
  \bibinfo{pages}{4559} (\bibinfo{year}{1970}).

\bibitem[{\citenamefont{Smit}(1955)}]{Smit1955}
\bibinfo{author}{\bibfnamefont{J.}~\bibnamefont{Smit}},
  \bibinfo{journal}{physica (Amsterdam)} \textbf{\bibinfo{volume}{21}},
  \bibinfo{pages}{877} (\bibinfo{year}{1955}), 
  \bibinfo{journal}{physica (Amsterdam)} \textbf{\bibinfo{volume}{24}},
  \bibinfo{pages}{39} (\bibinfo{year}{1958}).


\bibitem[{\citenamefont{Karplus and Luttinger}(1954)}]{Karplus1954}
\bibinfo{author}{\bibfnamefont{R.}~\bibnamefont{Karplus}} \bibnamefont{and}
  \bibinfo{author}{\bibfnamefont{J.}~\bibnamefont{Luttinger}},
  \bibinfo{journal}{Phys.\ Rev.} \textbf{\bibinfo{volume}{95}},
  \bibinfo{pages}{1154} (\bibinfo{year}{1954}).

\bibitem[{\citenamefont{Onoda and Nagaosa}(2003)}]{Onoda2003b}
\bibinfo{author}{\bibfnamefont{S.}~\bibnamefont{Onoda}} \bibnamefont{and}
  \bibinfo{author}{\bibfnamefont{N.}~\bibnamefont{Nagaosa}},
  \bibinfo{journal}{\prl} \textbf{\bibinfo{volume}{90}},
  \bibinfo{pages}{196602} (\bibinfo{year}{2003}).

\bibitem[{\citenamefont{Bhat et~al.}(2007)\citenamefont{Bhat, Kr\"{a}mer,
  Cooper, and Holland}}]{Bhat2007}
\bibinfo{author}{\bibfnamefont{R.}~\bibnamefont{Bhat}, {\it et al}},
  \bibinfo{journal}{\pra} \textbf{\bibinfo{volume}{76}},
  \bibinfo{pages}{043601} (\bibinfo{year}{2007}).

\bibitem[{\citenamefont{Viefers}(2008)}]{Viefers2008}
\bibinfo{author}{\bibfnamefont{S.}~\bibnamefont{Viefers}},
  \bibinfo{journal}{J.\ Phys.\ Condensed matters}
  \textbf{\bibinfo{volume}{20}}, \bibinfo{pages}{123202}
  (\bibinfo{year}{2008}).

\bibitem[{\citenamefont{Mandel et~al.}(2003)\citenamefont{Mandel, Greiner,
  Widera, Rom, H\"ansch, and Bloch}}]{Mandel2003}
\bibinfo{author}{\bibfnamefont{O.}~\bibnamefont{Mandel}, {\it et al}},
  \bibinfo{journal}{\prl} \textbf{\bibinfo{volume}{91}},
  \bibinfo{pages}{010407} (\bibinfo{year}{2003}).

\bibitem[{\citenamefont{Liu et~al.}(2007)\citenamefont{Liu, Liu, Kwek, and
  Oh}}]{Liu2007}
\bibinfo{author}{\bibfnamefont{X.-J.} \bibnamefont{Liu}, {\it et al}},
  \bibinfo{journal}{\prl} \textbf{\bibinfo{volume}{98}},
  \bibinfo{pages}{026602} (\bibinfo{year}{2007}).

\bibitem[{\citenamefont{Bliokh and Bliokh}(2006)}]{Bliokh2006}
\bibinfo{author}{\bibfnamefont{K.~Y.} \bibnamefont{Bliokh}} \bibnamefont{and}
  \bibinfo{author}{\bibfnamefont{Y.~P.} \bibnamefont{Bliokh}},
  \bibinfo{journal}{\prl} \textbf{\bibinfo{volume}{96}}, \bibinfo{eid}{073903}
  (\bibinfo{year}{2006}).

\bibitem[{\citenamefont{Zhu et~al.}(2006)\citenamefont{Zhu, Fu, Wu, Zhang, and
  Duan}}]{Zhu2006}
\bibinfo{author}{\bibfnamefont{S.-L.} \bibnamefont{Zhu}, {\it et al}},
  \bibinfo{journal}{\prl} \textbf{\bibinfo{volume}{97}}, \bibinfo{eid}{240401}
  (\bibinfo{year}{2006}).

\bibitem[{\citenamefont{Dudarev et~al.}(2004)\citenamefont{Dudarev, Diener,
  Carusotto, and Niu}}]{Dudarev2004}
\bibinfo{author}{\bibfnamefont{A.~M.} \bibnamefont{Dudarev}, {\it et al}},
  \bibinfo{journal}{Phys.\ Rev.\ Lett.} \textbf{\bibinfo{volume}{92}},
  \bibinfo{pages}{153005} (\bibinfo{year}{2004}).

\bibitem[{\citenamefont{Ruseckas et~al.}(2005)\citenamefont{Ruseckas,
  Juzeliunas, Ohberg, and Fleischhauer}}]{Ruseckas2005}
\bibinfo{author}{\bibfnamefont{J.}~\bibnamefont{Ruseckas}, {\it et al}},
  \bibinfo{journal}{\prl} \textbf{\bibinfo{volume}{95}},
  \bibinfo{pages}{010404} (\bibinfo{year}{2005}).

\bibitem[\citenamefont{Fort\'agh and Zimmermann}(2007)]{Fortagh2007}
  \bibinfo{author}{\bibnamefont{J.}~\bibnamefont{Fort\'agh}}
  \bibnamefont{and}
  \bibinfo{author}{\bibfnamefont{C.}~\bibnamefont{Zimmermann}},
  \bibinfo{journal}{Rev.\ Mod.\ Phys.} \textbf{\bibinfo{volume}{79}},
  \bibinfo{pages}{235} (\bibinfo{year}{2007}).
\bibitem{Nielsch2001}
  K. Nielsch {\it et al}, Appl. Phys. Lett. {\bf 79}, 1360 (2001)
\bibitem[{\citenamefont{Aharonov and Stern}(1992)}]{Aharonov1992}
\bibinfo{author}{\bibfnamefont{Y.}~\bibnamefont{Aharonov}} \bibnamefont{and}
  \bibinfo{author}{\bibfnamefont{A.}~\bibnamefont{Stern}},
  \bibinfo{journal}{Phys.\ Rev.\ Lett.} \textbf{\bibinfo{volume}{69}},
  \bibinfo{pages}{3593} (\bibinfo{year}{1992}).
\bibitem[{\citenamefont{Bruno et~al.}(2004)\citenamefont{Bruno, Dugaev, and
  Taillefumier}}]{Bruno2004}
\bibinfo{author}{\bibfnamefont{P.}~\bibnamefont{Bruno}, {\it et al}},
  \bibinfo{journal}{Phys. Rev. Lett.} \textbf{\bibinfo{volume}{93}},
  \bibinfo{pages}{96806} (\bibinfo{year}{2004}).

\bibitem[{\citenamefont{Ohmi and Machida}(1998)}]{Ohmi1998}
\bibinfo{author}{\bibfnamefont{T.}~\bibnamefont{Ohmi}} \bibnamefont{and}
  \bibinfo{author}{\bibfnamefont{K.}~\bibnamefont{Machida}},
  \bibinfo{journal}{J.\ Phys.\ Soc.\ Jpn.} \textbf{\bibinfo{volume}{67}},
  \bibinfo{pages}{1822} (\bibinfo{year}{1998}).

\bibitem[{\citenamefont{Ho}(1998)}]{Ho1998}
\bibinfo{author}{\bibfnamefont{T.-L.} \bibnamefont{Ho}},
  \bibinfo{journal}{Phys.\ Rev.\ Lett.} \textbf{\bibinfo{volume}{81}},
  \bibinfo{pages}{742} (\bibinfo{year}{1998}).

\bibitem[{\citenamefont{Castin and Dum}(1999)}]{Castin1999}
\bibinfo{author}{\bibfnamefont{Y.}~\bibnamefont{Castin}} \bibnamefont{and}
  \bibinfo{author}{\bibfnamefont{R.}~\bibnamefont{Dum}},
  \bibinfo{journal}{Eur.\ Phys.\ j.\ D} \textbf{\bibinfo{volume}{7}},
  \bibinfo{pages}{399} (\bibinfo{year}{1999}).

\bibitem[{\citenamefont{Bohm et~al.}(2003)\citenamefont{Bohm, Mostafazadeh,
  Koizumi, niu, and Zwanziger}}]{Bohm2003}
\bibinfo{author}{\bibfnamefont{A.}~\bibnamefont{Bohm}, {\it et al}},
  \emph{\bibinfo{title}{The Geometric Phase In quantum systems}}
  (\bibinfo{publisher}{Springer}, \bibinfo{year}{2003}), \bibinfo{note}{iSSN
  0172-5998}.

\bibitem[{not()}]{note}
\bibinfo{note}{The s-wave scaterring lengths for Rubidium are $a_0=101.8\ a_B$
  and $a_2=100.4\ a_B$ where $a_B$ is the Bohr radius\cite{Kempen2002}}.

\bibitem[{\citenamefont{van Kempen~{\it et al.}}(2002)}]{Kempen2002}
\bibinfo{author}{\bibfnamefont{E.}~\bibnamefont{van Kempen~{\it et al.}}},
  \bibinfo{journal}{\prl} \textbf{\bibinfo{volume}{88}},
  \bibinfo{pages}{093201} (\bibinfo{year}{2002}).

\end{thebibliography}
\end{document}